**Concurrent kilovoltage CBCT imaging and megavoltage beam delivery: Suppression of cross-scatter with 2D antiscatter grids and grid-based scatter sampling**

**Authors:** Farhang Bayat[a], Mohamed Elsayed Eldib[a], Brian Kavanagh[a], Moyed Miften[a], Cem Altunbas[a]

[a] Department of Radiation Oncology, University of Colorado School of Medicine, Aurora, CO 80045, USA.

**Corresponding Authors:**

1. Farhang Bayat

    Email address: farhang.bayat@cuanschutz.edu

    Postal Address: 1665 Aurora Court, Suite 1032 | MS F-706 | Aurora, CO 80045

2. Cem Altunbas

    Email address: cem.altunbas@cuanschutz.edu

    Postal Address: 1665 Aurora Court, Suite 1032 | MS F-706 | Aurora, CO 80045

**Abstract**

The concept of using kilovoltage (kV) and megavoltage (MV) beams concurrently has potential applications in cone beam computed tomography (CBCT) guided radiation therapy, such as single breath hold scans, metal artifact reduction, and simultaneous imaging during MV treatment delivery. However, MV cross-scatter generated during MV beam delivery degrades CBCT image quality. To address this, a 2D antiscatter grid and cross scatter correction method were investigated. a 3D printed, tungsten 2D antiscatter grid prototype was utilized to reduce MV cross-scatter fluence in kV projections during concurrent MV beam delivery. Remaining cross-scatter was corrected by using the 2D grid itself as a cross-scatter intensity sampling device, referred as Grid-based Scatter Sampling. To test this approach, kV CBCT acquisitions were performed while delivering 6 and 10 MV beams, mimicking high dose rate treatment delivery scenarios. MV cross-scatter suppression performance of the proposed approach was evaluated in projections and CBCT images of phantoms. 2D grid reduced the intensity of MV cross-scatter in kV projections by a factor of 3 on the average. Remaining MV cross-scatter estimated by Grid-based Scatter Sampling was within 7% of measured reference intensity values. CBCT image quality was improved substantially during concurrent kV-MV beam delivery. Median Hounsfield Unit (HU) inaccuracy was up to 191 HU without our methods, and it was reduced to 3 HU with our 2D grid and scatter correction approach. Our methods provided a factor of 2-6 improvement in contrast-to-ratio. Results indicate that our approach can successfully minimize the effects of high energy cross-scatter in concurrent kV CBCT imaging and megavoltage treatment delivery.

**1. Introduction**

In image-guided radiation therapy (IGRT), 2D or 3D kV image guidance has been temporally sequenced with respect to MV treatment beam, such that targets are localized either before treatment delivery or intermittently during treatment delivery. The concept of using kV and MV beams concurrently has potential advantages over temporally sequenced kV and MV beam delivery, and variety of novel applications have been proposed in recent years.

One of the promising applications of concurrent kV-MV beam delivery is in breath-hold CBCT, where MV and kV projections are acquired concurrently to reduce CBCT scan trajectory and duration by half. It has been shown that such kV-MV imaging scheme could reduce scan time down to a mere 15 seconds which could enable single breath-hold CBCT (Blessing *et al.*, 2010) (Wertz *et al.*, 2010). Furthermore, concurrent kV-MV beam delivery could be used for simultaneous 3D imaging during MV treatment delivery to improve intrafraction target localization and to calculate delivered dose (Ling *et al.*, 2011; Iramina *et al.*, 2020b; Iramina *et al.*, 2021; Boylan *et al.*, 2012; Iramina *et al.*, 2020a). Likewise, tomosynthesis-like fast 3D imaging approaches have been investigated by acquiring limited-angle kV and MV projections concurrently (Ren *et al.*, 2014). Another application of concurrent kV-MV projection acquisition is in metal artifact reduction, which exploits relative immunity of MV projections to metal-induced photon starvation artifacts (Wu *et al.*, 2014; Li *et al.*, 2013; Altunbas *et al.*, 2015).

However, concurrent kV-MV beam delivery results in MV cross-scattered x-rays impinging on the kV flat panel detector (FPD) which deteriorates the image quality. Several solutions have been proposed to address this problem. One suggests triggering kV and MV beam pulses in an alternating sequence, such that MV beam is off while kV beam is on and vice versa (Ling *et al.*, 2011). While this approach practically eliminates MV cross-scatter in kV projections, it reduces kV projection acquisition rate by 50% and may increase kV imaging duration. Moreover, such kV-MV pulse synchronization may cause MV beam interruptions at high dose rates typically used in Stereotactic Body Radiation Therapy (SBRT). Another approach is to reduce kV beam pulse rate, such that some of the kV FPD projections are acquired while only MV beam is on, and kV beam is off. This method allows measurement of MV cross-scatter in projections acquired without kV beam, and subsequent correction of MV cross-scatter intensity (Van Herk *et al.*, 2011). This approach negates MV beam interruptions but reduces kV image acquisition rate and cannot suppress noise due to MV cross-scatter. Model-based MV cross-scatter estimation methods have also been investigated to correct cross-scatter in kV projections(Iramina *et al.*, 2020b; Boylan *et al.*, 2012). However, these approaches may suffer from the discrepancy in modeled and actual clinical imaging conditions, leading to misestimation of cross-scatter intensity. Furthermore, noise due to cross-scatter is not suppressed in such scatter correction approaches.

In this work, we propose a new, two-stage approach to address MV cross-scatter problem. In the first stage, we implement a 2D antiscatter grid developed for kV CBCT imaging(Alexeev *et al.*, 2018; Altunbas *et al.*, 2017), to reject MV cross-scattered x-rays. This way, Hounsfield Unit (HU) loss can be partially recovered, and stochastic noise due to cross-scatter can be reduced. In the second stage, we implement a scatter correction method to correct MV cross-scatter that is not stopped by the grid. This method, referred to as Grid-based Scatter Sampling (GSS) (Altunbas *et al.*, 2021; Yu *et al.*, 2020), utilizes 2D grid as a scatter measurement device, and corrects both kV scatter and MV cross-scatter simultaneously in projections. GSS method can further reduce HU loss, thereby improving HU accuracy and reducing image artifacts. We performed experiments to investigate the utility of 2D grid and GSS method in the context of concurrent kV CBCT imaging and MV beam delivery during SBRT-like radiation treatment scenarios.

Scatter suppression methods used in this work, namely 2D antiscatter grids and GSS, were previously investigated in the context of kV scatter suppression (Altunbas *et al.*, 2021; Yu *et al.*, 2020). However, the utility of these methods has not been investigated in the context of MV cross-scatter suppression. Therefore, the novelty of the present work is in the robust mitigation of MV cross scatter problem in concurrent kV and MV beam delivery, by using clinically viable kV and MV beam delivery methods. To the best of our knowledge, a similar solution for MV cross-scatter suppression method has not been proposed. Our work, as explained in the following sections, allows MV cross scatter suppression while delivering kV and MV beams without pulse synchronization. Therefore, concurrent kV imaging and MV beam delivery can be achieved without interrupting the MV beam and without compromising kV image acquisition rate.

## 2. Method

### 2.1 Experiment Setup

To assess the effect of MV cross-scatter, two sets of imaging experiments were done, one with and other without MV beam delivery while acquiring kV CBCT scans using a Varian TrueBeam linac. To evaluate the effect of 2D grid on MV cross-scatter, a focused tungsten 2D antiscatter grid prototype was developed with a grid ratio of 12, grid pitch of 2 mm, and septal thickness of 0.1 mm, and it was installed on the FPD (Altunbas *et al.*, 2019) after removing the default (1D) antiscatter grid. For comparison, experiments were also conducted without 2D grid, but with default TrueBeam antiscatter grid in place (Altunbas *et al.*, 2017).

kV CBCT scans were performed in offset detector geometry (i.e., half fan geometry) by using the clinical pelvis protocol parameters; 900 projections were acquired at 125 kVp and 1080 mAs per scan with bow tie filter and 0.9 mm titanium foil in place. Since mAs setting and bowtie filter affect the primary signal intensity and hence the relative contribution of MV cross-scatter in projection signal, a subset of imaging experiments was repeated without bowtie filter and at 450 mAs per scan. Detector pixel size during acquisitions was 0.388×0.388 mm². Experiments were conducted by using head and pelvis sized electron density phantoms, as well as anatomically more realistic thorax and pelvis phantoms. All images were subsequently flat-field corrected and reconstructed using filtered backprojection(Biguri *et al.*, 2016).

kV CBCT imaging during SBRT delivery was emulated by setting the MV field size to 3x3 cm² and delivering 1200 Monitor Units (MU) at a rate of 1200 MU/min during kV CBCT acquisition, corresponding to 1.33 MU per kV projection. To evaluate the effect of cross-scatter intensity on image quality, the imaging protocol was repeated by increasing the field size to 10x10 cm² and reducing MU to 200 delivered at 200 MU/min (for clarity, delivered Monitor Units was assumed to be equal to delivered MV dose in cGy in the rest of the text). The effect of beam energy was evaluated by using 6 and 10 MV flattening filter free beams.

## 2.2. Grid-Based Scatter Sampling (GSS) Method

Our method employs 2D antiscatter grid as a residual scatter measurement device, which measures and corrects the intensity of residual scatter in projections (Altunbas *et al.*, 2021; Yu *et al.*, 2020). A brief explanation of the method follows.

When a 2D grid is placed on the detector, pixels located underneath the grid septa receive less x-ray fluence than the pixels located within grid holes (Fig. 1a). Ideally, this effect could be modeled and corrected by getting a second set of projections without the object (referred as flood projections) to correct such a signal reduction due to 2D grid's septal shadows in projections (Fig. 1b).

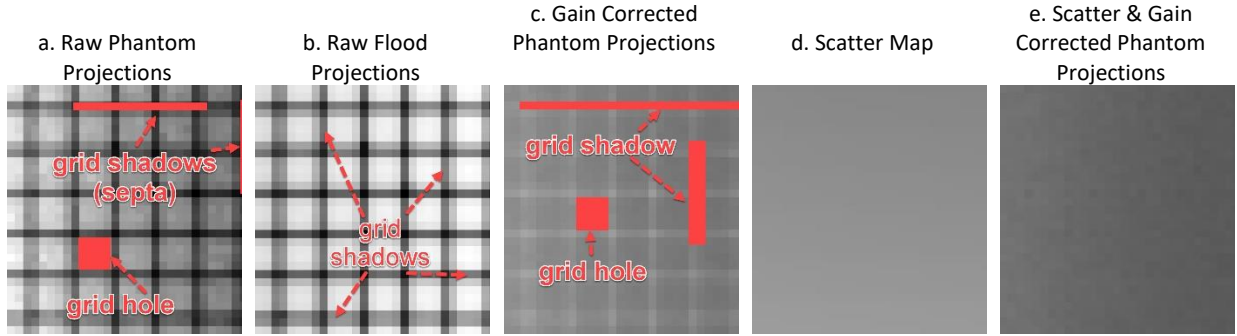

*Figure 1: Analysis of Grid Effect on flood, phantom and Gain Corrected Projections Before and After GSS*

This correction is referred as gain map (*GM*), and formulated as

$$GM(x,y) = \frac{C}{F(x,y)}$$

where C is an arbitrary normalization constant, and $F(x,y)$ is the flood projection. Thus, septal shadows in a raw projection can be compensated by multiplication with Gain Maps. However, when residual scatter is present in a CBCT scan, pixels within the grid shadows show a higher intensity than the pixels in a grid hole after gain map correction (Fig. 1c). This effect is due to the additive residual scatter signal in CBCT projections, whereas gain map is a multiplicative correction that is generated from scatter free flood projections.

In the GSS method, such a signal difference between grid shadows and grid holes is exploited to measure the residual scatter intensity. When residual scatter, $S$, is present in projections, hyperintense signal intensity pattern in Fig.1c changes as a function of $S$. Assuming

$S$ is piecewise uniform in pixels residing both in grid shadows and grid holes in a small neighborhood of pixels (small neighborhood is a 7 × 7 pixel region, corresponding to an area of 2.7 × 2.7 mm$^2$), $S$ can be calculated as(Altunbas *et al.*, 2021; Yu *et al.*, 2020),

$$S(x_1, y_1) = \frac{d(x_1, y_1)}{GM_{grid}(x_1, y_1) - GM_{hole}(x_2, y_2)} \quad (1)$$

Where $x_1$ and $y_1$ are for pixels in grid shadows and $x_2$ and $y_2$ are for pixels in grid holes, $d$ is the signal difference in grid shadows and holes in a small neighborhood, $GM_{grid}$ and $GM_{hole}$ are the values of gain maps in grid septal shadows and holes, respectively.

Thus, two major assumptions in the GSS method are 1) primary kV signal intensity is reduced by 2D grid's grid shadows, 2) scatter intensity is the same, or uniform, in pixels residing in grid shadows and grid holes, in a small neighborhood of pixels. Previously, these assumptions in the GSS method were shown to be acceptable in suppressing kV scatter(Altunbas *et al.*, 2021; Yu *et al.*, 2020). These assumptions are considered valid when MV cross-scatter is present.

Using the above formulations, we could thus estimate scatter in grid shadows and use interpolation to find residual scatter values in each detector pixel. We finally subtracted the calculated scatter (Fig. 1d) from projections to achieve scatter corrected projections (Fig. 1e) and proceeded to image reconstruction.

It is important to note that GSS method is fundamentally different than widely adopted beam stop methods(Altunbas, 2014) in the following ways: 1) In beam stop based methods, beam stops cast a large shadow on multiple pixels, and signal in the shadow is equal to the scatter intensity. Whereas, in our method, the grid wall thickness is only a fraction of a detector pixel size, and signal in grid shadows is a mixture of both primary and scatter signals. As a result, scatter intensity in 2D grid's shadow cannot be measured directly, as in a beam stop method. In the GSS method, the scatter intensity was measured via the change in the signal intensity introduced by the 2D grid's shadows. 2) In beam stop methods, a beam stop array is placed between the patient and the x-ray source. In our case, the 2D grid, i.e., the scatter measurement device, is placed between the patient and the detector.

### 2.3. Measures of Comparison

**Projection domain evaluations:** MV cross-scatter rejection and residual cross-scatter estimation performance were evaluated in projections.

To achieve this, first MV cross-scatter intensity in kV projections was measured during concurrent kV-MV beam delivery, which served as baseline, or reference, to assess the performance of our methods. For a given phantom configuration, two CBCT data sets were acquired, one with kV only and the other with concurrent kV-MV beams. Projections in these two data sets were paired by matching the source angle in each pair. Subsequently, subtraction of kV only projection from kV-MV projection in each pair yielded the *reference* MV cross-scatter intensity as a function of gantry angle. At each gantry angle, mean MV cross-scatter intensity was calculated in a 100x100 pixel wide region of interest (ROI) that was centered at the piercing point (i.e., projected location of focal spot in a projection). This process was repeated with default TrueBeam 1D antiscatter grid and with 2D grid to evaluate the effect of grid type on MV cross-scatter intensity. Ratio of cross-scatter intensities with default 1D grid and with 2D grid yielded scatter rejection performance of 2D grid.

Likewise, Scatter Estimation Error (SEE) of the GSS method was calculated for each projection pair as below,

$$SEE = 100 \times \frac{|\text{mean}(\text{GSS based MV scatter} - \text{reference scatter})|}{\text{mean}(\text{reference MV cross scatter})}$$

**Image domain evaluations:** HU loss represents how HU values degrade in kV-MV images due to MV cross-scatter when compared to kV-only images. HU loss in kV-MV images was evaluated for default 1D grid, 2D grid and 2D grid + GSS configurations with their respective kV only counterparts. Specifically, we introduce $\Delta HU_{method}$ which represents the average absolute HU difference between any method's output for kV-MV and their output for their kV only reconstruction. The formula is as follows:

$$\Delta HU_{method} = |HU_{kV\ Only, method} - HU_{kV+MV, method}| \qquad (3)$$

This approach assures that HU loss is solely due to MV cross-scatter for any given scatter suppression method evaluated.

To evaluate HU loss in each image set, eight ROIs were selected in the contrast objects of the electron density phantoms, and subsequently, mean HU loss was calculated for each ROI. Median and range of HU loss was reported for 8 ROIs in each image set.

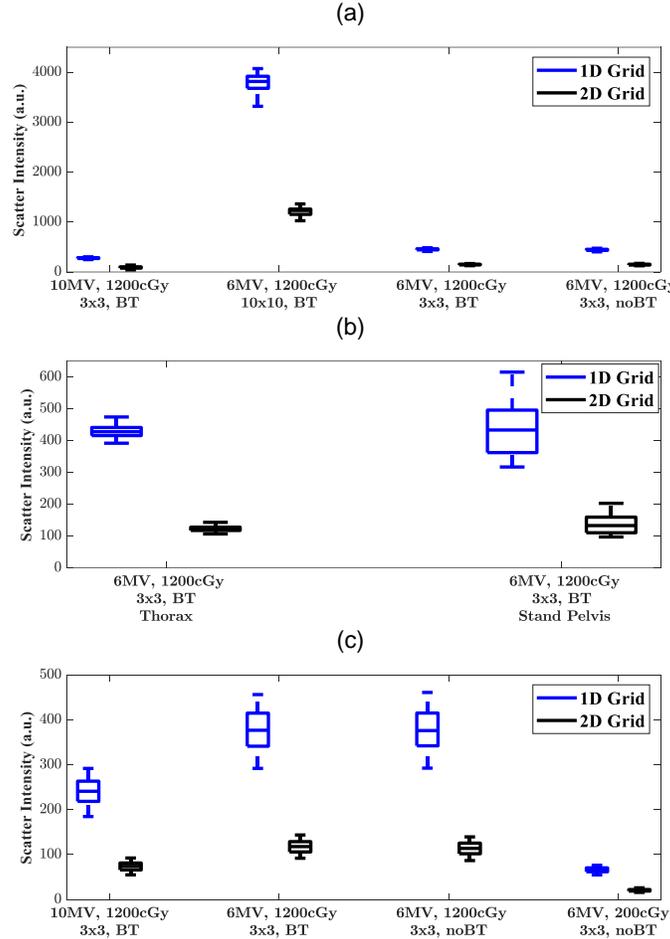

Fig. 2: MV cross-scatter intensity comparison between 1D and 2D Grids for (a) head sized electron density phantom, (b) pelvis sized electron density phantom, and (c) thorax and pelvis phantoms

Relative Contrast-to-noise Ratio improvement factor (kCNR) was calculated using the same ROIs in the electron density phantom, and indicates the change in CNR in kV-MV images in comparison with CNR values in the absence of 2D grid (Altunbas et al., 2019).

Finally, we introduced Identity Function Profile (IFP) as a graphical representation of HU loss. Ideally, if all scatter due to MV-cross scatter were suppressed, pixel-by-pixel HU values would be the same for kV only and kV-MV reconstructions. Thus, if we were to plot a histogram between each corresponding pixel, the optimal scatter suppression method would result in an identity function. Hence, we could measure the optimality of any suggested scatter suppression method by evaluating how close to an identity function their profile resides.

Attenuation coefficient to HU conversion was calculated by measuring attenuation coefficients in water equivalent background section of the head-sized electron density phantom. This measurement was done using 2D grid only CBCT images acquired at 125 kVp, without the GSS correction method. Subsequently, the same water attenuation coefficient was used for HU conversion in 1D grid and 2D grid + GSS images. This process was repeated for CBCT scans acquired with BT filter. A separate HU conversion factor was not used for each grid type, because our goal was to show the relative change in attenuation coefficients in 3 different scatter mitigation configurations we have investigated.

## 3. Results

### 3.1. Projection Domain

A comparison of average scatter values for different protocols is presented in Fig. 2. (a), (b) and (c). When compared to the default 1D Grid in TrueBeam, 2D Grid provided significant MV cross-scatter rejection, and reduced MV cross-scatter intensity by a factor of 3.19±0.15 on the average across all protocols.

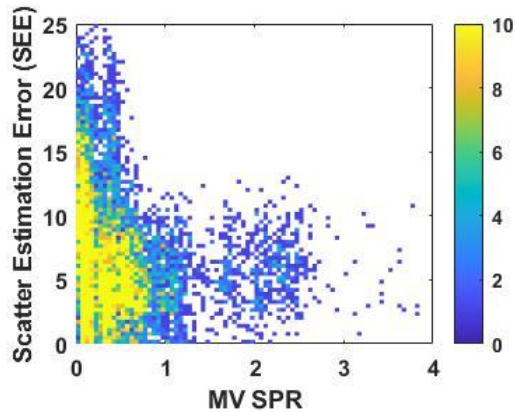

*Fig. 3: MV cross-scatter estimation error of the GSS method versus scatter to primary ratio in each CBCT projection, across all protocols.*

Fig. 3 shows the SEE for each imaging protocol versus their respective MV cross-scatter to Primary Ratio (SPR) in the same ROI. As can be seen, SEE is relatively close to 0 in all scenarios with a mean of 6.97± 3.76 %. while cross-scatter estimation performance of the GSS method appears to be lower when MV SPR is below 0.5. This was in part attributed to inaccuracies in measuring reference (or ground truth) MV cross-scatter values. This issue was further elaborated in the Discussions section.

### 3.2. Reconstruction Domain

When 2D Grid was combined with GSS residual scatter correction, HU loss, averaged across all protocols, was further reduced to 6 HU (Fig. 4). Qualitatively, difference images (Fig. 4) also show that the effect of MV cross-scatter on HU accuracy was minimized when 2D grid + GSS approach was used. HU loss in 6 MV ad 10 MV beam deliveries were comparable, 12.5HU vs 3.5HU respectively, when 2D grid + GSS was used, indicating applicability of our method to different MV beam energies. When residual MV cross-scatter was present, ring artifacts were induced in images, particularly in images acquired with the 2D grid (Fig. 4). These artifacts were suppressed after cross-scatter correction with GSS method.

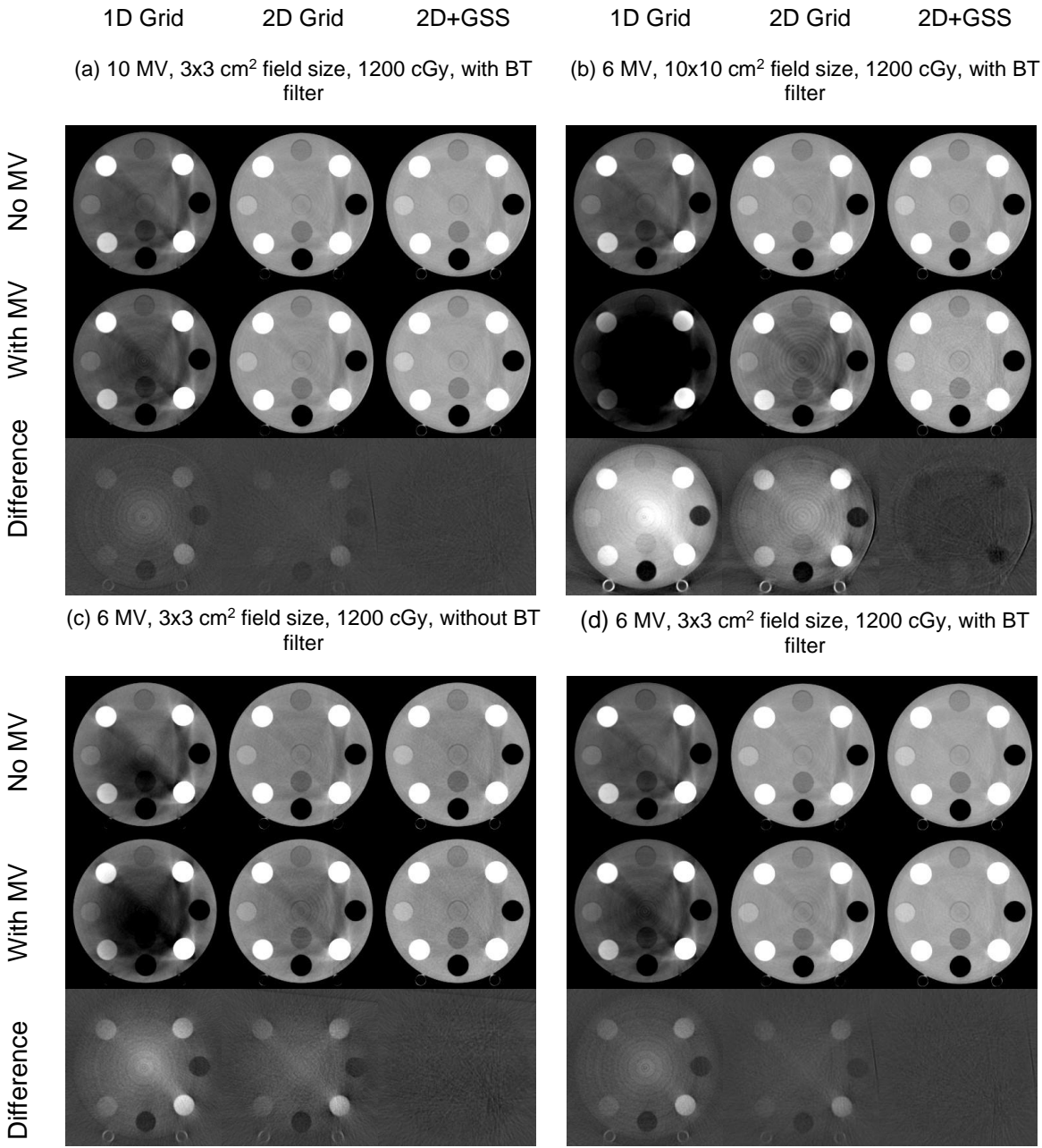

Fig. 4: Reconstructions for head-sized electron density phantom protocols, the window levels are [-250 250] for No MV and With MV and [-100 300] for Difference

Fig. 4 shows the effect of MV cross-scatter on image quality in head-sized phantoms. The difference in images with and without MV beam demonstrates the HU degradation qualitatively. As expected, median HU loss -in the absence of a 2D grid- was severe; HU loss was 191 for the case of 1200 cGy dose delivery with 6MV beam and 10x10 cm2 field size, while using the CBCT scan protocol with bow tie (BT) filter (Fig. 4b). MV field size and kV mAs were the two major factors that affected the HU loss (as observed by comparing the pair of Fig. 4a and Fig. 4b). When 2D grid was in use, HU loss, averaged across all protocols, was reduced from 71 HU to 36 HU.

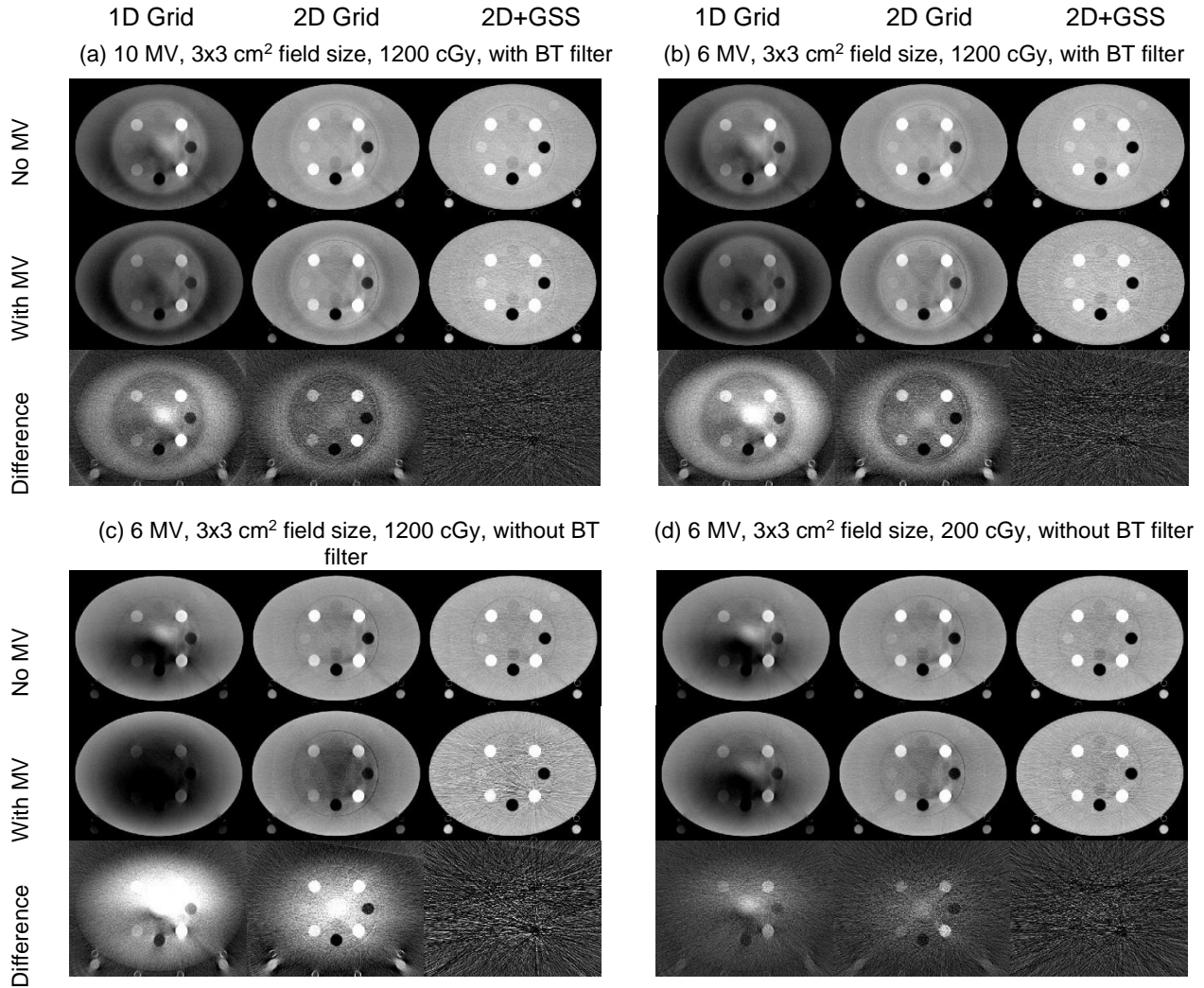

*Fig. 5: Reconstructions of pelvis-sized electron density phantom protocols. HU window ranges are [-450 250] for No MV and With MV and [-50 200] for Difference.*

A similar trend was observed in pelvis-sized phantom images (Fig. 5). Due to larger phantom size, MV SPR was higher, which yielded even larger HU degradation. The use of 2D grid reduced HU loss across all protocols from 97 HU to 64 HU. When 2D grid was combined with GSS, mean HU loss was further reduced to 16.8 HU. While reduction of MV beam dose from 1200 to 200 cGy led to proportionally lower HU loss, our method promptly recovered HU values in both cases; mean HU loss was reduced from 158 HU to 39 HU and from 39 HU to 16 HU in 2D Grid only and 2D grid + GSS imaging protocols, respectively. Images acquired without bow tie filter (Figs. 5c and 5d) had lower kV imaging dose, and therefore, the intensity and detrimental effects of MV cross-scatter were relatively large in these image sets. While GSS method restored the HU values to a large extent, increased noise due to MV cross-scatter was visible in images. Increase in image noise was less pronounced in images acquired with bow tie filter and imaging dose (Figs. 5a and 5b).

HU loss as a function of imaging protocols is gathered in Fig. 6 which includes box plots of HU loss in (a) head and (b) pelvis sized phantoms. Each box plot consists of one box and two whiskers where the line within the box is median; upper and lower edges of the box correspond to 75th and 25th percentile and upper and lower whiskers correspond to maximum and minimum data points. Across all imaging protocols investigated, 2D grid + GSS approach consistently led to substantial reduction in HU loss. Increasing the MV field size from 3x3 to 10x10 cm$^2$ caused the highest increase in MV-cross scatter fluence among head-sized phantoms, and hence, increased HU loss up to 340HU in some ROIs without 2D grid. With 2D Grid and 2D grid + GSS, median HU loss was reduced to 96HU and 16.5HU on the average, respectively. In images acquired without BT filter, HU loss was also substantially larger, as well as visually apparent in CBCT images (Figs. 4c, 5c, 5d).

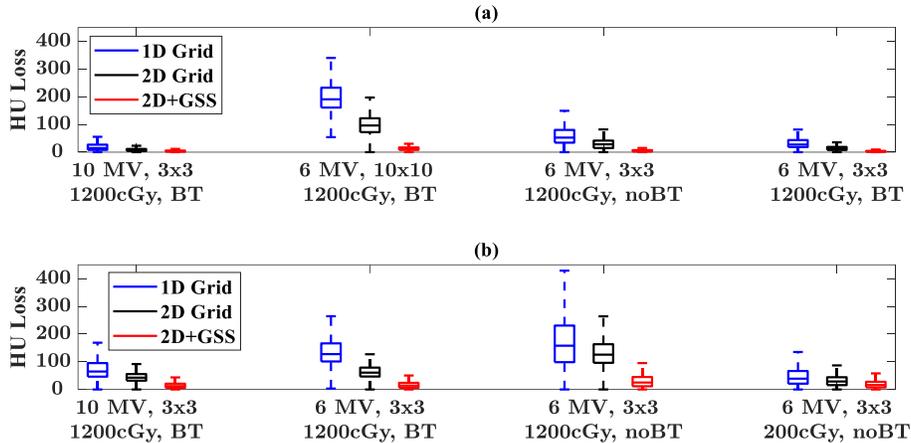

Fig. 6: HU Loss in (a) Head-sized (b) Pelvis-sized electron density phantoms as a function of imaging protocol.

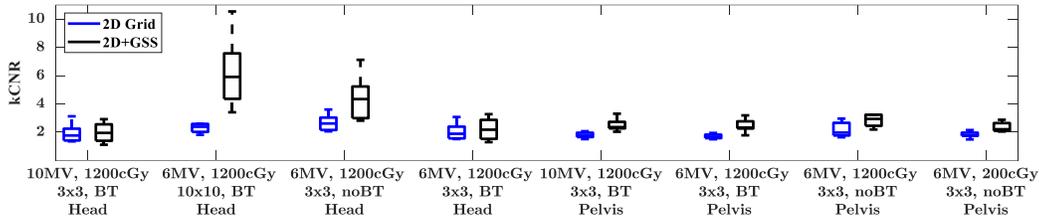

Fig. 7: kCNR for Head-sized and Pelvis-sized electron density phantoms as a function of imaging protocol. kCNR represents CNR improvement factor with respect to images acquired without 2D grid

As in indicated in Fig. 7, 2D grid rejects a large amount of scatter which improved median CNR by a factor of 2 across all imaging protocols. Addition of GSS further improved CNR and resulted in a factor of 3 improvement in median CNR values across all protocols in comparison with the original 1D Grid. Increase in CNR with GSS method was largely due to reduction of shading artifacts, and associated reduction in calculated noise in ROIs.

Next, anatomically realistic pelvis and thorax phantom images are shown in Figs. 8a and 8b, respectively. In pelvis phantom, median HU loss among ROIs reached 150HU, when 2D grid was not used. The loss of HU accuracy was also evident in the difference images. With 2D grid, median HU loss decreased to 14-20 HU range. When 2D grid was combined with GSS residual scatter correction, HU loss was further reduced to 5.7 HU. Similarly for thorax, HU losses were up to 200HU. When 2D grid was in use, HU loss decreased to 45HU. When 2D grid was combined with GSS residual scatter correction, median HU loss was reduced to 7 HU. Difference images qualitatively show the high degree of agreement in HU values of kV-only and kV-MV acquisitions, when 2D grid + GSS approach was used.

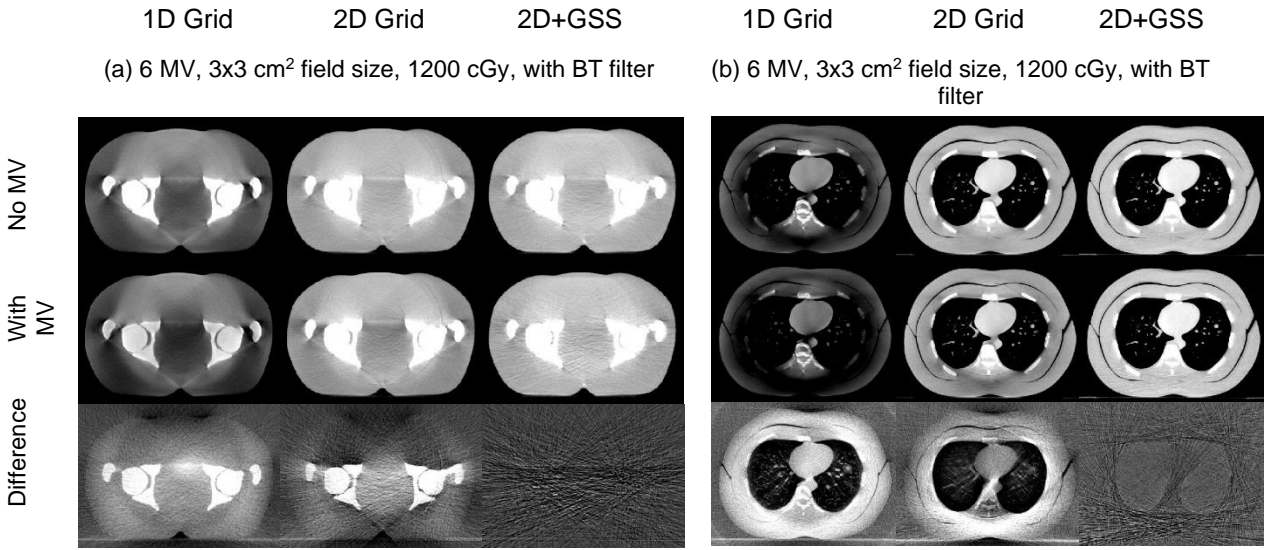

*Fig. 8: Reconstructions of (a) Pelvis (b) Thorax phantoms. HU window range, pelvis phantom: [-400 200] for CBCT images and [-50 150] for difference. Thorax phantom: [-450 150] for CBCT images and [-50 100] for difference.*

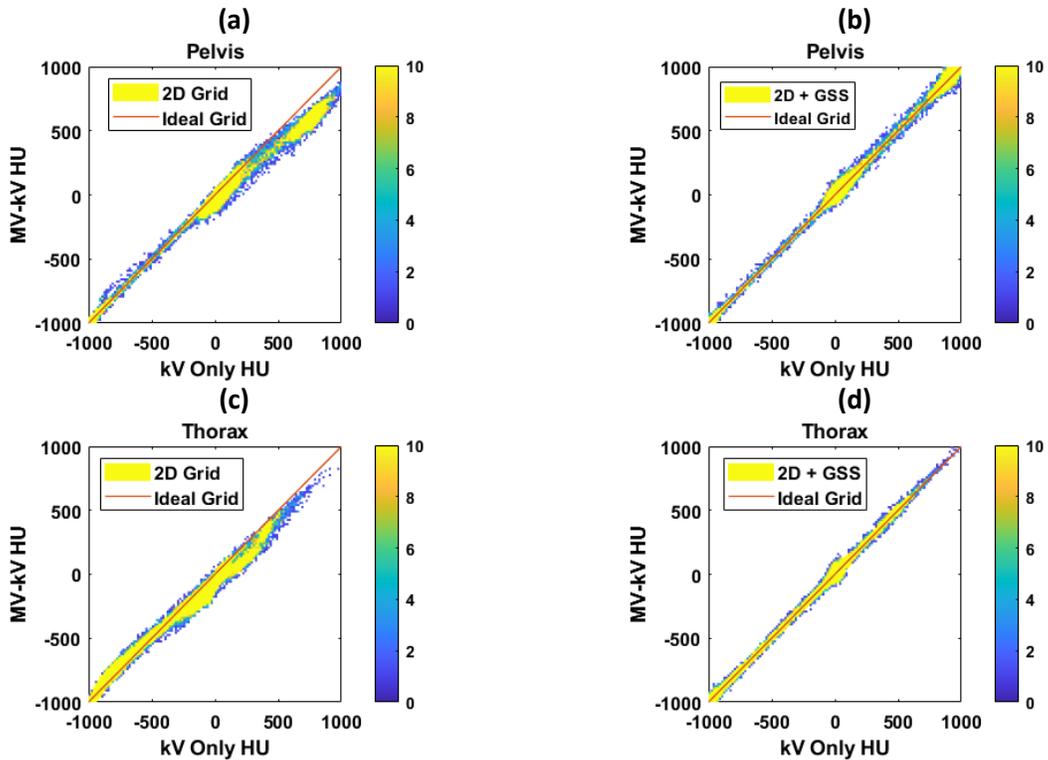

*Fig. 9: HU correlation between kV-MV and kV only CBCT scans for Pelvis (a, b) and Thorax (c, d) phantoms. HU correlations were plotted for 2D grid only and 2D+GSS configurations.*

Finally, IFPs of using 2D grid without and with GSS over pelvis and thorax are presented in Fig. 9. While 2D grid generates an IFP close to the ideal value (identity function) for both pelvis and thorax, there is a subset of pixels where a noticeable divergence from the ideal profile is evident, indicating that there are pixels whose HU values are not restored reasonably well (Figs. 9a and

9c). These pixels often correspond to high density regions, such as bony anatomy, where MV cross-scatter to primary ratio was higher in their projections.

## 4. Discussion

In this work, we introduced a novel and robust method of rejecting and correcting MV cross scatter in kV projections. A 2D antiscatter grid developed for kV imaging rejected MV cross-scatter with a mean energy of 300 keV(Taylor *et al.*, 1999) as indicated in Figs. 2, 6 and 7, thereby improving CT number accuracy and CNR. Specifically, CNR improvement is a significant advantage of cross-scatter rejection with 2D antiscatter grids, which cannot be achieved with scatter correction methods(Rührnschopf and Klingenbeck, 2011; Ruhrnschopf and Klingenbeck, 2011). However, MV-cross scatter fluence was not fully rejected by the 2D grid, and effects of residual MV cross-scatter incident on the FPD were still evident in CBCT images. As demonstrated in this work, our Grid-based Scatter Sampling (GSS) method was efficient in correcting residual cross-scatter (Figs. 6 and 7), such that CT number accuracy in concurrent kV-MV beam delivery was comparable to kV-only CBCT images. HU accuracy and CNR improvement were comparable for 6 and 10 MV beams, indicating that beam energy plays a minor role in mitigating the effects of cross-scatter.

In addition to MV cross-scatter, MV x-rays due to linac head leakage were also part of the contaminant MV image signal in kV projections. Since our methods corrected all sources of contaminant x-rays regardless of their origin, effects of head leakage on kV projections were also expected to be reduced by our methods. However, performance evaluation of our methods in suppressing the effects of head leakage was not studied separately, an area of potential future research.

In contrast to kV-only scatter, the effects of MV cross-scatter strongly depend on the kV imaging dose, or kV primary signal. This is because, the fraction of MV-cross scatter in image signal is larger at lower kV doses, and vice versa. This effect was demonstrated in Figs. 5b and 5c; images acquired with bow tie filter have twice the kV primary signal intensity (when compared to no bow tie filter scans), and MV cross-scatter to primary ratio was halved. Therefore, HU loss in images acquired with bow tie filter was less (Fig. 6b). While GSS method restores HU accuracy to a large extent, CNR loss may not be recovered, as evidenced by relatively noisy appearance in images acquired without bow tie filter (Fig.5c). Thus, CNR loss and HU loss due to MV cross-scatter can be further reduced by using higher kV imaging dose, when feasible.

Besides HU and CNR loss, another drawback of residual cross-scatter is the induction of ring artifacts in reconstructed images. Such ring artifacts are particularly visible in Fig. 4b, in images acquired with 2D grid. This is mostly due to reduced efficacy of flat-field correction in the presence of residual scatter, and hence, suboptimal correction of grid's septal shadows (Altunbas *et al.*, 2021; Yu *et al.*, 2020). Since GSS suppresses residual scatter effectively, it also reduces the ring artifacts caused by cross-scatter. Similar artifacts were also observed when default 1D grid was in place, as seen in Figs 4a and 4b.

MV cross-scatter intensity estimated by the GSS method was in good agreement with the measured reference values, when MV SPR was above 0.5. However, the accuracy of GSS method appeared to deteriorate at lower MV SPR values. Our observations indicated that this issue stemmed from two different sources. First, reference MV cross-scatter intensity values were measured by subtracting kV-only projections from kV-MV projections, which was considered the ground truth. In kV-MV acquisitions, kV and MV beams were triggered asynchronously as in a clinical kV-MV beam delivery scenario. Such asynchronous triggering caused delivery of MV pulses during kV detector readout phase, which manifests itself as detector row-to-row variations in MV cross-scatter intensity, and stripe artifacts in projections. Such high spatial frequency signal variations might not be fully accounted by the GSS method. This issue was apparent when MV beam pulses were triggered at lower frequencies, such as during 6 MV beam delivery with 200 MU (Fig. 5). Second, it was assumed that the image signal difference between kV-only and kV-MV projections was equivalent to MV cross-scatter scatter intensity. This approach assumes that the kV image signal is identical in a given kV and kV-MV projection pair acquired at the same source angle. However, source angles in kV and kV-MV projection pairs in a CBCT acquisition can be different from each other by 0.2-0.3 degrees, causing different kV beam attenuation paths and kV image signal intensities in these two data sets. Moreover, kV tube output, and detector response cannot be kept identical in separate kV-only and kV-MV acquisitions. Such variations in kV image signal intensity ultimately affect the accuracy of MV cross-scatter measurement. Such inaccuracies were accentuated when MV SPR is low, or in other words, majority of kV-MV projection signal is composed of kV-only signals. Problems listed above are inherent limitations of measuring reference MV cross-scatter intensity, which, in return, affects the performance assessment of the GSS method in low MV SPR imaging conditions.

Most of the data in this work was acquired using 1200 MU and 3x3 cm$^2$ field size, to emulate beam delivery conditions in SBRT treatments. Based on our clinical experience, multi-leaf collimator (MLC) aperture areas vary largely during SBRT delivery, which would affect MV cross-scatter intensity. In addition, delivered MU per treatment beam often covers a wide range, depending on the treatment site, dose constraints, treatment planning techniques and dose prescription. Therefore, a more accurate assessment of MV cross-scatter suppression can be done in the future by using variety of clinical SBRT delivery scenarios and SBRT treatment plans.

One of the concerns in using 2D antiscatter grids is the reduced primary signal, and its potential adverse effects on CNR. While antiscatter grids reduce primary fluence incident on the detector and increase noise, scatter rejection provided by the antiscatter grid improves contrast. Therefore, the overall change in CNR depends on the interplay between the CNR degradation caused by primary reduction and CNR improvement due to scatter rejection provided by the antiscatter grid. The 2D grid used in this study has a primary transmission fraction of 85%, whereas the 1D grid has primary transmission fraction of 70% (Altunbas *et al.*, 2019; Altunbas *et al.*, 2017). As a result, CNR degradation due to primary beam attenuation by the antiscatter grid is less with the 2D grid. Moreover, the 2D grid provides a factor of 3.3 to 7.3 (depending on the phantom thickness) better scatter rejection than 1D grid, and associated CNR improvement due to scatter rejection is higher with 2D grid (Altunbas *et al.*, 2017; Altunbas *et al.*, 2019). Thus, due to higher primary transmission and efficient scatter rejection, CNR improvement by 2D grid can be achieved without increasing the imaging dose in head and pelvis sized phantoms as investigated in prior studies(Park *et al.*, 2021). In this work, further CNR improvement with 2D grids was realized when kV and MV beams were used concurrently. This is due to substantially better MV cross-scatter intensity rejection performance of 2D grids and reduction of image artifacts, when compared to 1D grids.

The 2D grid prototype used in this study has a substantially larger grid pitch than the grid pitch of the conventional 1D grid (2 mm versus 0.167 mm). There are two major reasons for using such a large grid pitch in the 2D grid. First, large grid pitch reduces the footprint of the tungsten septa on the detector, and thus improves primary transmission. Such large grid pitches are feasible to fabricate due to self-supporting structure of the 2D wall array. Second, our method for correcting any leftover scatter with the GSS method -as explained in Section 2.2- requires clear definition of grid holes and grid wall shadows in projections. Thus, grid pitch is required to be larger than the pixel pitch.

Although MV cross-scatter rejection by 0.1 mm thick tungsten walls may seem counter intuitive, it can be justified by analyzing the energy spectrum of the MV cross-scatter (Taylor *et*

*al.*, 1999). The average energy of MV cross scatter goes down as a function of exit angle (exit angle refers to the angle between the primary MV beam direction and the detector used to measure MV cross scatter). While average energy of 6-10 MV primary beam is about 2 MeV, average energy of cross scatter at 90-degree exit angle is about 240-300 keV. Another important detail is the quantum efficiency of the detector. While the scatter rejection performance of the 2D grid is reduced at higher energies, such high energy scattered x-rays are less likely to interact in the detector and contribute to image signal. Such energy dependent response of the detector to MV cross-scatter can be potentially investigated in simulations in a future study. Moreover, MV cross scatter rejection performance of our 2D grid was measured with respect to the 1D grid, which is composed of lead lamellae. Even though the linear attenuation coefficient of tungsten is reduced at 240 keV, similar behavior is also expected for the lead lamellae of the 1D grid. As a result, it is reasonable to expect better MV cross-scatter suppression performance from the 2D grid when compared to 1D grid.

Finally, we note how CBCT images acquired with 1D grid and bow tie filter show severe shading artifacts in the periphery of the phantom (Fig. 4 and 5). This is largely caused by the increased scatter-to-primary ratio in the periphery of the projections due to bow tie filter [6,1]. In clinical CBCT images, such shading artifacts are substantially less due to scatter correction. Whereas, in this work, additional scatter correction methods were not employed with the 1D grid.

## 5. Conclusion

Overall, a series of similar observations on different phantoms and imaging protocols indicate the strength of using the proposed 2D antiscatter grid and residual scatter correction method in mitigating the effects of MV cross-scatter in concurrent kV CBCT and MV beam delivery. HU loss due to MV cross-scatter was recovered to a large extent, and factor of 2 to 3 improvement in CNR was observed in concurrent kV CBCT scans and MV beam delivery. Unlike previously suggested methods, our methods do not require MV beam interruption or reduction in kV image acquisition rate, which makes them suitable for kV imaging during high dose rate MV therapy delivery, such as SBRT, and for fast kV-MV image acquisition protocols, such as breath-hold CBCT.

While proposed methods were evaluated in the context of image guided external beam radiation therapy, our methods can potentially be used in other applications such as imaging during high dose rate brachytherapy, and beyond radiation therapy, such as high x-ray energy industrial imaging and security imaging applications.

**Acknowledgements**

This work was funded in part by grants from NIH/NCI R21CA198462 and R01CA245270.